\documentclass[fleqn,10pt]{wlscirep}
\usepackage[utf8]{inputenc}
\usepackage[T1]{fontenc}

\usepackage{epsf, latexsym, color, epsfig}
\usepackage{amssymb, amsmath}
\usepackage{times}

\newcommand{\ket}[1]{| #1 \rangle}

\newcommand{\pra}{{\it Phys. Rev. A~}}
\newcommand{\prb}{{\it Phys. Rev. B~}}
\newcommand{\prl}{{\it Phys. Rev. Lett.~}}
\newcommand{\rmp}{{\it Rev. Mod. Phys.~}}

\newcommand\h{{\cal H}}
\newcommand{\etal}{{  \textit{et al.}}}

\newcommand{\ignore}[1]{}

\newcommand{\be}{\begin{equation}}
\newcommand{\ee}{\end{equation}}
\newcommand{\ba}{\begin{eqnarray}}
\newcommand{\ea}{\end{eqnarray}}

\def\CC{{\rm\kern.24em \vrule width.04em height1.46ex depth-.07ex
    \kern-.30em C}}
\def\P{{\rm I\kern-.25em P}}

\def\RR{{\rm
         \vrule width.04em height1.58ex depth-.0ex
         \kern-.04em R}}

\def\bbbc{{\mathchoice {\setbox0=\hbox{$\displaystyle\rm C$}\hbox{\hbox
to0pt{\kern0.4\wd0\vrule height0.9\ht0\hss}\box0}}
{\setbox0=\hbox{$\textstyle\rm C$}\hbox{\hbox
to0pt{\kern0.4\wd0\vrule height0.9\ht0\hss}\box0}}
{\setbox0=\hbox{$\scriptstyle\rm C$}\hbox{\hbox
to0pt{\kern0.4\wd0\vrule height0.9\ht0\hss}\box0}}
{\setbox0=\hbox{$\scriptscriptstyle\rm C$}\hbox{\hbox
to0pt{\kern0.4\wd0\vrule height0.9\ht0\hss}\box0}}}}

\def\bbbq{{\mathchoice {\setbox0=\hbox{$\displaystyle\rm Q$}\hbox{\raise
0.15\ht0\hbox to0pt{\kern0.4\wd0\vrule height0.8\ht0\hss}\box0}}
{\setbox0=\hbox{$\textstyle\rm Q$}\hbox{\raise
0.15\ht0\hbox to0pt{\kern0.4\wd0\vrule height0.8\ht0\hss}\box0}}
{\setbox0=\hbox{$\scriptstyle\rm Q$}\hbox{\raise
0.15\ht0\hbox to0pt{\kern0.4\wd0\vrule height0.7\ht0\hss}\box0}}
{\setbox0=\hbox{$\scriptscriptstyle\rm Q$}\hbox{\raise
0.15\ht0\hbox to0pt{\kern0.4\wd0\vrule height0.7\ht0\hss}\box0}}}}

\def\bbbt{{\mathchoice {\setbox0=\hbox{$\displaystyle\rm
T$}\hbox{\hbox to0pt{\kern0.3\wd0\vrule height0.9\ht0\hss}\box0}}
{\setbox0=\hbox{$\textstyle\rm T$}\hbox{\hbox
to0pt{\kern0.3\wd0\vrule height0.9\ht0\hss}\box0}}
{\setbox6=\hbox{$\scriptstyle\rm T$}\hbox{\hbox
to0pt{\kern8.3\wd0\vrule height0.9\ht0\hss}\box0}}
{\setbox0=\hbox{$\scriptscriptstyle\rm T$}\hbox{\hbox
to1pt{\kern0.3\wd1\vrule height0.9\ht0\hss}\box0}}}}

\def\bbbz{{\mathchoice {\hbox{$\sf\textstyle Z\kern-0.4em Z$}}
{\hbox{$\sf\textstyle Z\kern-0.4em Z$}}
{\hbox{$\sf\scriptstyle Z\kern-0.3em Z$}}
{\hbox{$\sf\scriptscriptstyle Z\kern-0.2em Z$}}}}

\begin{document}

\title{Cyclic permutations for qudits in $d$ dimensions}

\author[1]{Tudor-Alexandru Isdrail\u a}
\author[2]{Cristian Kusko}
\author[1,*]{Radu Ionicioiu}

\affil[1]{Horia Hulubei National Institute of Physics and Nuclear Engineering, Bucharest--M\u agurele 077125, Romania}
\affil[2]{National Institute for Research and Development in Microtechnologies IMT, Bucharest 077190, Romania}

\affil[*]{r.ionicioiu@theory.nipne.ro}

\begin{abstract}
One of the main challenges in quantum technologies is the ability to control individual quantum systems. This task becomes increasingly difficult as the dimension of the system grows. Here we propose a general setup for cyclic permutations $X_d$ in $d$ dimensions, a major primitive for constructing arbitrary qudit gates. Using orbital angular momentum states as a qudit, the simplest implementation of the $X_d$ gate in $d$ dimensions requires a single quantum sorter $S_d$ and two spiral phase plates. We then extend this construction to a generalised $X_d(p)$ gate to perform a cyclic permutation of a set of $d$, equally spaced values $\{ \ket{\ell_0}, \ket{\ell_0+p},\ldots, \ket{\ell_0+(d-1)p} \} \mapsto \{ \ket{\ell_0+p}, \ket {\ell_0+2p},\ldots, \ket{\ell_0} \}$. We find compact implementations for the generalised $X_d(p)$ gate in both Michelson (one sorter $S_d$, two spiral phase plates) and Mach-Zehnder configurations (two sorters $S_d$, two spiral phase plates). Remarkably, the number of spiral phase plates is independent of the qudit dimension $d$. Our architecture for $X_d$ and generalised $X_d(p)$ gate will enable complex quantum algorithms for qudits, for example quantum protocols using photonic OAM states.
\end{abstract}

\maketitle

\section*{Introduction}

All successful technologies are based on harnessing a specific resource, such as energy, electricity or information. The ability to generate, control, transform and ultimately, find useful applications for quantum resources, is central to the development of quantum technologies \cite{feynman, ddv, 2qrev, qkd_review}.

Controlling the simplest quantum systems, the qubit, is relatively straightforward \cite{Barenco, mike_ike}. We can achieve complete control over the 2-dimensional Hilbert space of a qubit with rotations generated by Pauli matrices $X$ and $Z$. The natural next step is to go to $d$-dimensional systems, or qudits. In this case we have the generalised Pauli matrices $X_d$ and $Z_d$. Progressing in this direction, we need to find physical implementations for qudits, together with the experimental ability to control them.

Orbital angular momentum (OAM) is one of the most used implementations for photonic qudits. Photon states $\ket{\ell}$ carry an OAM of $\ell\hbar$, where $\ell=0, \pm1, \pm2,\ldots $ is a theoretically unbounded integer. OAM states have a helical phase front, with $\ell\ne 0$ corresponding to the number of helices.

Photonic OAM states have been used in entanglement generation \cite{mair, fickler, krenn1} and alignment-free quantum key distribution \cite{align_free, vallone_oam}. Thus OAM is attractive since it allows us to use a larger alphabet to transmit quantum information with a single photon. However, without the appropriate tools, a larger alphabet for encoding information has only a limited functionality. This brings us to the problem of how to implement efficiently the generalised Pauli operators $X_d$ and $Z_d$ for qudits \cite{bloch_qudit}.

For photonic OAM states, $Z_d$ can be implemented with Dove prisms. An open question is how to implement a cyclic permutation $X_d$ for any dimension $d$. Experimentally, cyclic $X_d$ gates for OAM states have been realised only for $d=4$ \cite{cyclic1} and $d=5$ \cite{cyclic2}.

In this article we propose a general scheme to perform cyclic permutations $X_d$ for any set of $d$ consecutive states. We then generalise it for cyclic permutations $X_d(p)$ of an arbitrary set of $d$, equally spaced states $\{ \ket{\ell_0}, \ket{\ell_0+p},\ldots, \ket{\ell_0+(d-1)p} \}$. For any dimension $d$, the minimal implementation of both $X_d$ and $X_d(p) $ requires a single sorter $S_d$ and two spiral phase plates (SPPs) \cite{Beijer, Allen, Oemr, Wang}. To arrive at this setup, we use quantum information methods and quantum network analysis. This approach has been employed previously to design a universal quantum sorter \cite{sorter} and spin measuring devices \cite{ri1, ri3}.

We focus on OAM encoded qudits, as several experimental tools are already available \cite{leach1, leach2, giovannini, berk, mirh, babazadeh, LG_sorter}. Nevertheless, our scheme can be extended in principle to other degrees of freedom as well.

\section*{Results}

\subsection*{Cyclic $X_d$ gate}

We now introduce our setup for performing the cyclic gate. Let $\h_d$ be the Hilbert space of a qudit, $\dim \h_d= d$ and let $\{\ket{j} \}_{j=0}^{d-1}$ be an orthonormal basis of $\h_d$. The generalised Pauli operators $X_d$, $Z_d$ are defined as:
\ba
X_d:\ \ &&\ket{j} \mapsto \ket{j\oplus 1}\\
Z_d:\ \ &&\ket{j} \mapsto \omega^j \ket{j}
\label{Xd}
\ea
with $\oplus$ addition mod $d$ and $\omega= e^{2\pi i/d}$ a root of unity of order $d$. The gate $X_d$ performs a cyclic permutation of the basis states, i.e., maps the set $\{ \ket{0}, \ket{1}, \ldots, \ket{d-1}\}$ to $\{ \ket{1}, \ket{2}, \ldots, \ket{0}\}$.

\begin{figure}
\centering
\includegraphics[width=0.4\textwidth]{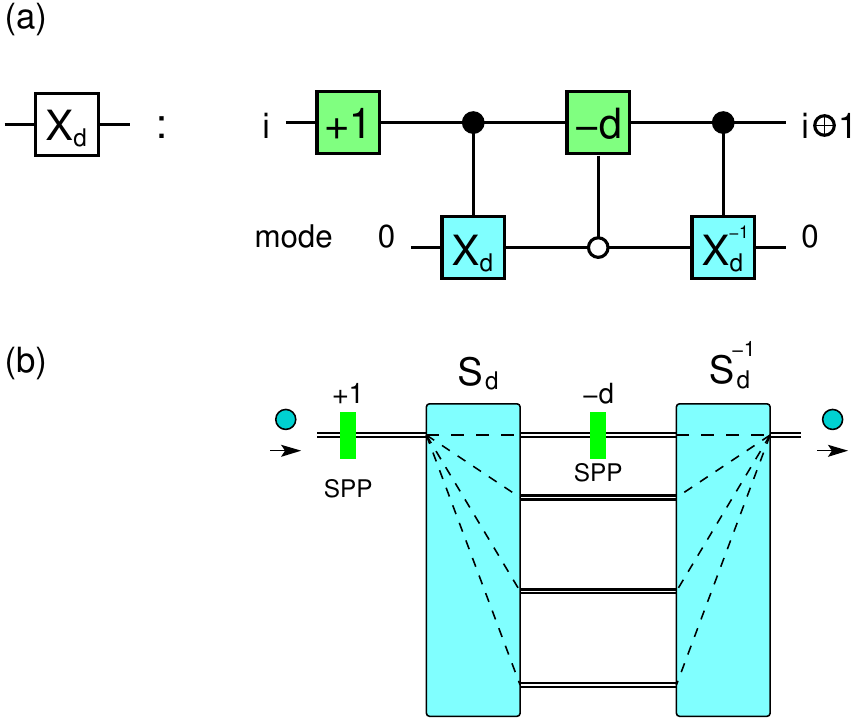}
\caption{Cyclic $X_d$ gate. The gate performs the transformation ${\ket{i} \mapsto \ket{i\oplus 1}}$ on a qudit. (a) Equivalent quantum network. Spatial modes are used as a qudit ancilla, which is factorised before and after the gate, i.e., starts and ends up in the state $\ket{0}_m$. There are two SPPs of order $+1$ and $-d$ (green) and two $C(X_d)$ gates (cyan). The $\mathrm{SPP}(-d)$ acts only on mode $0$ (open circle on the control mode qudit). (b) Implementation. Photons enter from the left and all OAM states are shifted by $\mathrm{SPP}(+1)$. The sorter $S_d$ redirects each state to the corresponding output. The state $\ket{d}_{OAM}$ exits on spatial mode $0$ and after the $\mathrm{SPP}(-d)$ becomes $\ket{0}_{OAM}$; all other OAM states $\ket{j}, j\ne d$, are left invariant. Finally, all states enter the inverse sorter $S_d^{-1}$ and end up in the same spatial mode $\ket{0}_m$.}
\label{Xd_MZI}
\end{figure}

Our scheme for the $X_d$ gate is shown in Fig.\ref{Xd_MZI}. The main element of our proposal is a $d$-dimensional sorter $S_d$ introduced in Ref.~\cite{sorter, sorter1}. A quantum sorter $S_d$ is a device which directs an incoming particle into different outputs (i.e., sorts) according to the value of an internal degree of freedom $\Sigma$. In the following we take $\Sigma$ to be orbital angular momentum (OAM). Nevertheless, the setup is general and can be implemented for other variables as well, like wavelength \cite{sorter} or radial quantum number \cite{radial_s1, radial_s2}.

\begin{figure}
\centering
\includegraphics[width=0.4\textwidth]{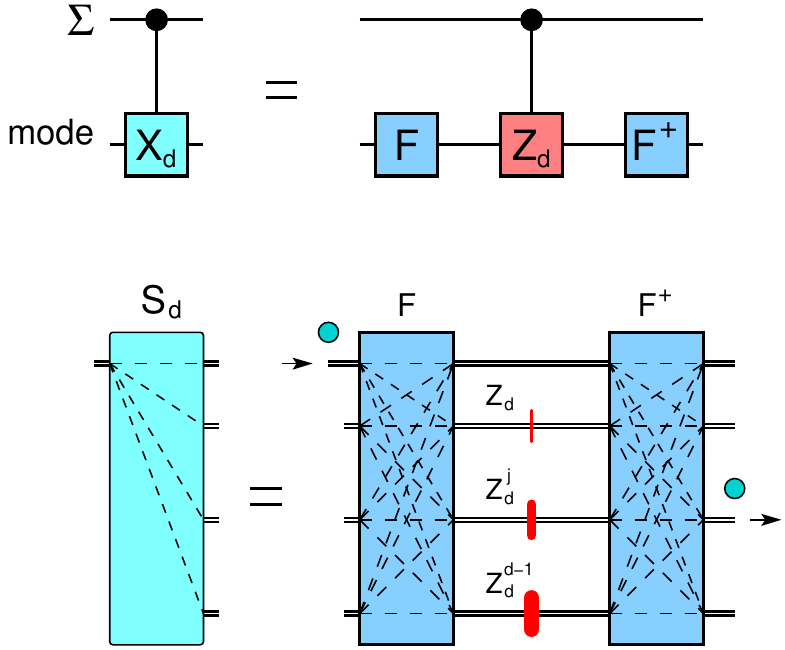}
\caption{Universal quantum sorter $S_d$. Top: equivalent quantum network. The controlled-$X_d$ gate $C(X_d)$ is decomposed on Fourier gates $F, F^\dag$ and a $C(Z_d)$ gate. Bottom: implementation as a multimode interferometer with path-dependent phase shifts $Z_d^j$ acting on the variable to be sorted $\Sigma$.}
\label{sorter}
\end{figure}

The quantum sorter $S_d$ is formally equivalent to a controlled-$X_d$ gate between the degree of freedom we want to sort (OAM, $\Sigma$ etc) and spatial modes $m$, see Fig.~\ref{sorter}:
\be
S_d:= C(X_d):\ \ \ket{i}_{OAM} \ket{j}_m \mapsto \ket{i}_{OAM} \ket{j\oplus i}_m
\label{S_d}
\ee
where $\ket{i}_{OAM}, \ket{j}_m$ are OAM and mode qudits, respectively. Thus a photon in OAM state $\ket{i}$ incident on port (mode) 0 will exit on port $(i \bmod d)$ with unit probability.

Apart from the sorter $S_d$ another ingredient are spiral phase plates \cite{Beijer, Allen, Oemr, Wang} of order $n$. The action of the SPP on OAM states is:
\be
\mathrm{SPP}(n):\ \ \ket{i}_{OAM} \mapsto \ket{i+n}_{OAM}
\label{SPP}
\ee
with $n \in \mathbb{Z}$ integer. This transformation adds (or subtracts) $n$ units of OAM. Since this is normal addition, it shifts the whole $\mathbb{Z}$ axis by $n$ units.

We now discuss how the $X_d$ gate in Fig.~\ref{Xd_MZI} works. The first SPP adds $+1$ to all OAM states. Then the sorter $S_d$ directs each OAM state $\ket{i}_{OAM}$ to the corresponding output $\ket{i\bmod d}_m$, eq.~\eqref{S_d}. Since sorting on modes is done modulo $d$, the state $\ket{d}_{OAM}$ will exit on mode $0$. Consequently, only the state on mode $0$ needs to be shifted by $-d$; in terms of quantum networks, this is equivalent to a controlled-$\mathrm{SPP}(-d)$ gate, with the control on the mode $k=0$ (open circle on control qudit in Fig.\ref{Xd_MZI}). After this operation the states from all spatial modes are recombined on mode $0$ by the gate $C(X_d)^{-1}$, which is nothing else but a sorter run in reverse $S_d^{-1}$. This decouples the OAM and mode qudits, such that the final state is factorised and the photon always exits on mode 0 with unit probability. Thus the gate in Fig.~\ref{Xd_MZI} performs the following sequence: 
\ba
\nonumber \ket{i}_{OAM} \ket{0}_m &\stackrel{+1}{\longrightarrow}& \ket{i+1}_{OAM} \ket{0}_m \\
\nonumber &\stackrel{S_d}{\longrightarrow}& \ket{i+1}_{OAM} \ket{i\oplus 1}_m \\
\nonumber &\stackrel{-d^{[0]}}{\longrightarrow}& \ket{i\oplus 1}_{OAM} \ket{i\oplus 1}_m \\
&\stackrel{S_d^{-1}}{\longrightarrow}& \ket{i\oplus 1}_{OAM} \ket{0}_m
\ea

Since the ancilla is decoupled after the gate, an arbitrary superposition of OAM states transforms under the cyclic gate $X_d$ as
\be
\alpha_0\ket{0}+ \alpha_1 \ket{1}+ \ldots + \alpha_{d-1}\ket{d-1}\ \  \rightarrow\ \  \alpha_{d-1}\ket{0}+ \alpha_0\ket{1}+ \ldots+ \alpha_{d-2}\ket{d-1}
\ee
Consequently, our scheme preserves coherence and can be used in arbitrary quantum algorithms.

\subsection*{Resources}

Our implementation of the cyclic $X_d$ gate requires two sorters $S_d$ and two SPPs (of order $+1$ and $-d$, respectively). This result is important: the number of SPPs is constant (two), and thus independent of the qudit dimension $d$. A Michelson configuration simplifies this to a single sorter and two SPPs, see section below.

Each sorter $S_d$ requires $d-1$ phases $Z_d$ (on OAM) and two Fourier gates $F_d, F_d^\dag$ on spatial modes \cite{sorter}, Fig.~\ref{sorter}. Fourier gates for spatial modes \cite{barak, taki, tabia, kita} can be implemented in several ways. 

In integrated optics the Fourier gate is performed by a single slab coupler \cite{cincotti}, e.g., as used in arrayed waveguide gratings (AWG) \cite{WavLow}. Commercially available AWGs containing Fourier gates have tens to hundreds of spatial modes (channels).

In bulk optics the Fourier transform can be implemented with a pair of confocal lenses with waveguides attached \cite{cincotti}. Alternatively, the Fourier gate can be decomposed in $O(d^2)$ linear optics elements (beamsplitters and phase-shifters) \cite{tabia, reck, clements}. For $d=2^p, p\in \mathbb N$, the Fourier transform on $d$ modes can be implemented with $O(\tfrac{d}{2} \log d)$ linear optical elements (beamsplitters and phase-shifters) \cite{barak, crespi}. The resource scaling for the $X_d$ gate is summarised in Table \ref{t1}.

\begin{table}[h]
\begin{center}
\begin{tabular}{l||c|c|c}
& SPPs & Fourier $F$ & $Z_d$ phases \\
\hline
resources & 2 & 4, integrated optics & $2(d-1)$\\
& & $O(d^2)$, linear optics & \\
& & $O(\tfrac{d}{2} \log d)$, if $d=2^p$ & 
\end{tabular}
\end{center}
\caption{Resource scaling for the cyclic $X_d$ gate.}
\label{t1}
\end{table}

We now briefly discuss possible implementations. Given the scaling discussed above and the current advances in photonics, we expect to have sizes $d\sim 5-10$ in bulk optics and $d\sim 100$ in integrated photonics.

\subsection*{Generalised $X_d$ gate}

The previous setup implements a cyclic $X_d$ gate on the set $\{ \ket{0}, \ket{1}, \ldots, \ket{d-1}\}$. We now discuss two generalisations.

\begin{figure}[t]
\centering
\includegraphics[width=0.5\textwidth]{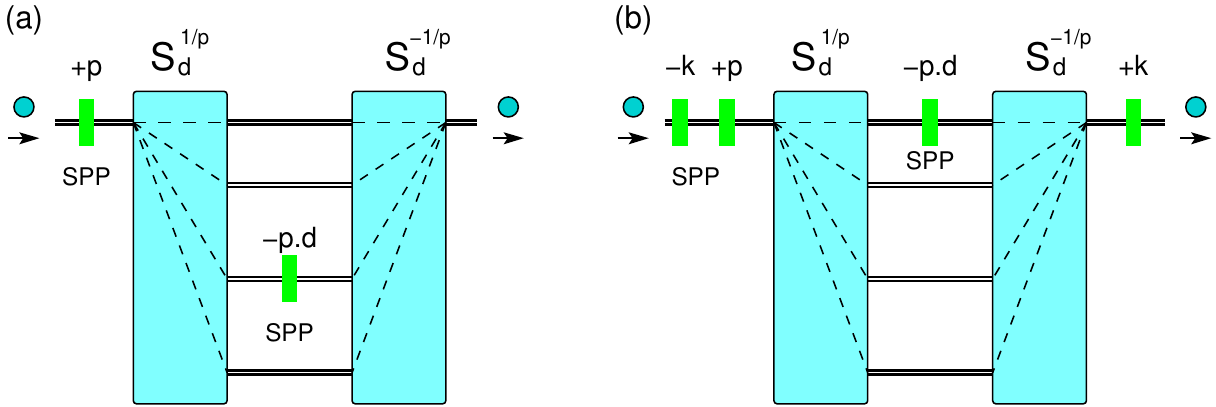}
\caption{Generalised cyclic gate $X_d(p)$. The gate performs a cyclic permutation on the set $\{ \ket{\ell_0}, \ket{\ell_0+p},\ldots, \ket{\ell_0+(d-1)p} \}$. There are two ways to correct for the initial state $\ell_0$: (a) by moving $\mathrm{SPP}(-pd)$ to spatial mode $k$, with $k= \ell_0 \bmod d$; (b) by inserting before and after the sorters two SPPs, $\mathrm{SPP}(-k)$ and $\mathrm{SPP}(k)$. Note that $0\le k< d$.}
\label{Xdp_MZI}
\end{figure}

\noindent {\em (i) Cyclic permutation for $d$ consecutive values.}
The first generalisation is to perform a cyclic $X_d$ gate on an arbitrary set of $d$ consecutive OAM values $\{ \ket{\ell_0}, \ket{\ell_0+1}, \ldots , \ket{\ell_0+d-1}\}$. From the general transformation of the sorter, we know that a state with OAM $\ell_0$, incident on spatial mode $0$, will exit on spatial mode $k= \ell_0 \bmod d$. Thus the only modification of the scheme in Fig.~\ref{Xd_MZI} is to move $\mathrm{SPP}(-d)$ from mode 0 to mode $k$, Fig.~\ref{Xdp_MZI}(a).

A second method to perform the cyclic $X_d$ gate on the set $\{ \ket{\ell_0}, \ket{\ell_0+1}, \ldots , \ket{\ell_0+d-1}\}$ is to first shift all OAM states with $-k$, then perform the usual $X_d$ gate, and finally shift back all states with $+k$, Fig.~\ref{Xdp_MZI}(b).

Importantly, the value of the shift $k$ is bounded by the dimension $d$ of the qudit and not by $\ell_0$, since $0\le k <d$. This is noteworthy -- although OAM with large values $\ell=10010$ have been experimentally prepared \cite{OAM10010}, SPPs with such large values are very difficult to manufacture. Therefore, for $\ell \gg d$ we need to shift the state only with a much smaller value ${(\ell_0 \bmod d)} < d$, irrespective of the magnitude of $\ell_0$. Thus the same device can be used to perform the cyclic $X_d$ gate on any set of $d$ consecutive OAM states. In this case the only change is the position of $\mathrm{SPP}(-d)$ (for the first method), or adding two extra $\mathrm{SPP}(\pm k)$ (for the second method).

\noindent {\em (ii) Cyclic permutation for $d$, equally spaced values.}
Let $p\in \mathbb{N}^+$ be a positive integer. We now show how to implement a cyclic permutation with step $p$
\be
\{ \ket{0}, \ket{p},\ldots, \ket{(d-1)p} \} \mapsto \{ \ket{p}, \ket {2p},\ldots, \ket{0} \}
\ee
We define the generalised gate $X_d(p)$
\be
X_d(p):\ \ \ket{j p} \mapsto \ket{(j\oplus 1) p}
\ee
with $j=0,\ldots, d-1$, $p\in \mathbb{N}^+$ and $\oplus$ addition mod $d$. The gate is characterised by two parameters: the qudit dimension $d$ and the step $p$ between two consecutive values; clearly $X_d(1)= X_d$.

The implementation of $X_d(p)$ uses the same architecture as before, but with a different sorter $S_d^{1/p}$: this directs each state $\ket{j p}_{OAM}$ on a separate mode $\ket{j}_m$. From the decomposition $S_d^{1/p}= F^\dag C(Z_d^{1/p}) F$, it follows that we can implement $S_d^{1/p}$ with a setup similar to Fig.~\ref{sorter}, but with different mode-dependent phases $Z_d^{j/p}$ inside the interferometer.

The setup for $X_d(p)$ is shown in Fig.~\ref{Xdp_MZI}. As before, $\mathrm{SPP}(p)$ first shift all OAM states with $p$. The sorter $S_d^{1/p}$ separates the states according to the OAM values, $\ket{j p}_{OAM} \ket{0}_m \mapsto \ket{j p}_{OAM} \ket{j}_m$. Then $\mathrm{SPP}(-pd)$ maps $\ket{dp}_{OAM} \mapsto \ket{0}_{OAM}$, after which the sorter $S_d^{-1/p}$ combines back all states on mode $\ket{0}_m$.

As before, we can implement a cyclic gate $X_d(p)$ on an arbitrary set of equally spaced OAM values $\{ \ket{\ell_0}, \ket{\ell_0+p},\ldots, \ket{\ell_0+(d-1)p} \}$ in two ways, Fig.~\ref{Xdp_MZI}:\\
(a) by moving $\mathrm{SPP}(-pd)$ on mode $k$, with $k= \ell_0 \bmod d$; or\\
(b) by using two $\mathrm{SPP}(\pm k)$ before and after the sorters.

To summarise, for any dimension $d$ and initial state $\ell_0$, the generalised gate $X_d(p)$ requires only two sorters $S_d^{1/p}, S_d^{-1/p}$ and two spiral phase plates $\mathrm{SPP}(p)$, $\mathrm{SPP}(-pd)$.

\subsection*{Simplification: Michelson setup}

\begin{figure}
\centering
\includegraphics[width=0.5\textwidth]{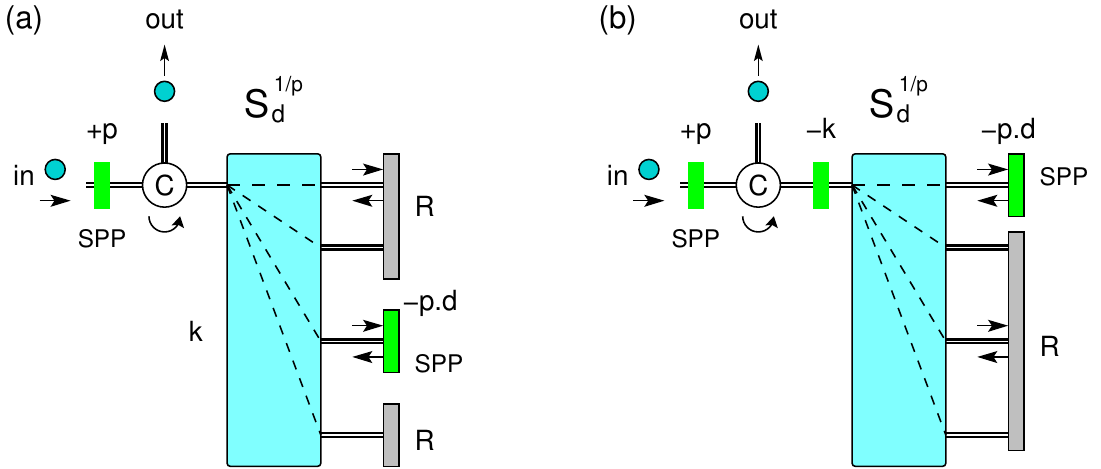}
\caption{Michelson interferometer configuration for $X_d(p)$ gate. This is the folded version of the Mach-Zehnder setup in Fig.\ref{Xdp_MZI}. The state $\ket{dp}_{OAM}$ exits on spatial mode $0$ and after a reflection on $\mathrm{SPP}(-pd)$ becomes $\ket{0}_{OAM}$. All other OAM states $\ket{jp}_{OAM}, j\ne d$, are left invariant by a double reflection on retro-reflectors $R$. Subsequently, all states re-enter the sorter from the opposite direction, thus performing $S_d^{-1/p}$, and end up in the same spatial mode $\ket{0}_m$. A circulator $C$ separates the output from the input. Cases (a) and (b) are equivalent to Fig.~\ref{Xdp_MZI} (a), (b), respectively.}
\label{Xdp_MI}
\end{figure}

We can further simplify our scheme if we use a Michelson instead of the Mach-Zehnder interferometer. In this case we need only one sorter $S_d^{1/p}$, Fig.~\ref{Xdp_MI}. The first part of the scheme is identical to the one discussed previously. The state $\ket{dp}_{OAM}$ exits on spatial mode $k$ and, after a reflection on $\mathrm{SPP}(-pd)$, becomes $\ket{0}_{OAM}$. All other OAM states undergo a double reflection on the retro-reflector $R$ and remain unchanged. Finally, all states re-enter the sorter from the opposite direction, thus performing $S_d^{-1/p}$, and end up in the same spatial mode $\ket{0}_m$. A circulator $C$ separates the output from the input. Note that a spiral phase plate acts as its inverse if the photon enters from the opposite direction; thus in Fig~\ref{Xdp_MI}(b) we need only a single $\mathrm{SPP}(-k)$.

\section*{Discussion}

The ability to control higher-dimensional quantum systems is essential for developing useful quantum technologies. Due to coherence constraints, efficiency will play a key role in the success of real-life quantum protocols. In this article we designed implementations for cyclic permutations $X_d$ in $d$ dimensions, one of the building blocks for constructing arbitrary single-qudit gates. The scheme is deterministic, works at single-particle level and can be applied to arbitrary superpositions of qudit states. Regarding the resource scaling, both $X_d$ and $X_d(p)$ gates require a linear number of phase-shifts $Z_d$. Remarkably, for any dimension $d$ the number of spiral phase plates SPPs is two, thus constant.

Although our focus has been on orbital angular momentum, the method is general and can be adapted to other degrees of freedom. This will require a sorter $S_d$ and shift gates (the equivalent of SPP) for the respective degree of freedom. Since a general scheme for a universal quantum sorter exists \cite{sorter}, a future challenge to implement the cyclic gate $X_d$ for a particular degree of freedom $\Sigma$ is to find appropriate implementations for shift gates $\ket{i}_\Sigma \mapsto \ket{i+n}_\Sigma$.

A possible application of the generalised cyclic permutation $X_d(p)$ is quantum communication, e.g., in QKD protocols with Fibonacci coding for key distribution \cite{FibSimon, FibZiwen}.

\noindent{\em Note added.} While finishing this article we became aware of another method for performing $X_d$ gates for OAM \cite{Xd_gao}.


\section*{Acknowledgements}

The authors acknowledge support from a grant of the Romanian Ministry of Research and Innovation, PCCDI-UEFISCDI, project number PN-III-P1-1.2-PCCDI-2017-0338/79PCCDI/2018, within PNCDI III. R.I.~acknowledges support from PN 18090101/2018.

\section*{Author contributions statement}

C.K. and R.I. proposed the theoretical method. T.A.I. and R.I. wrote the manuscript. All authors reviewed the manuscript.

\section*{Additional information}

\textbf{Competing interests:} The authors declare no competing interests.

\end{document}